\PassOptionsToPackage{dvipsnames}{xcolor}
\documentclass[sigconf, nonacm, screen]{acmart}

\AtBeginDocument{%
  }

\usepackage[all]{nowidow}
\usepackage{xcolor}
\usepackage{subcaption}
\usepackage{hyperref}
\usepackage{acronym}
\usepackage{siunitx}
\usepackage[capitalise,noabbrev]{cleveref}
\crefformat{section}{\S#2#1#3} %
\crefformat{subsection}{\S#2#1#3}
\crefformat{subsubsection}{\S#2#1#3}
\crefrangeformat{section}{\S#3#1#4 to~\S#5#2#6}
\crefrangeformat{subsection}{\S#3#1#4 to~\S#5#2#6}
\crefrangeformat{subsubsection}{\S#3#1#4 to~\S#5#2#6}

\begin{document}

\title[Same Function, Different CPU]{Same Function, Different CPU: Quantifying Hardware Heterogeneity in Commercial FaaS Platforms}

\author{Jonas Heisterberg}
\affiliation{%
    \institution{Technische Universit\"at Berlin}
    \city{Berlin}
    \country{Germany}}
\email{jh@3s.tu-berlin.de}
\orcid{0009-0004-4863-4018}

\author{Natalie Carl}
\affiliation{%
    \institution{Technische Universit\"at Berlin}
    \city{Berlin}
    \country{Germany}}
\email{nc@3s.tu-berlin.de}
\orcid{0009-0000-5991-9255}

\author{Trever Schirmer}
\affiliation{%
    \institution{Technische Universit\"at Berlin}
    \city{Berlin}
    \country{Germany}}
\email{ts@3s.tu-berlin.de}
\orcid{0000-0001-9277-3032}

\author{David Bermbach}
\affiliation{%
    \institution{Technische Universit\"at Berlin}
    \city{Berlin}
    \country{Germany}}
\email{db@3s.tu-berlin.de}
\orcid{0000-0002-7524-3256}

\renewcommand{\shortauthors}{Heisterberg et al.}

\begin{abstract}
Serverless computing, particularly Function-as-a-Service (FaaS), abstracts operational concerns such as scaling, scheduling, and resource management from users.
This abstraction, however, conceals a heterogeneous hardware fleet on which function execution environments are scheduled behind a unified function interface.
Consequently, otherwise identical invocations may run on different CPU types that differ in performance, a hardware lottery that users can neither see nor control.
Since the resulting variability is entangled with other sources, such as noisy neighbors or diurnal effects, its contribution to overall performance variability and how this contribution differs across providers remain unclear.

In this paper, we systematically study hardware heterogeneity and its impact on function performance across four widely used commercial FaaS platforms, covering regional, configurational, and temporal differences.
Using variance decomposition, we find that CPU heterogeneity alone explains \qtyrange{54}{97}{\percent} of the performance variance on AWS Lambda and Google Cloud Run Functions.
On Azure Functions, in contrast, we identify a platform-level mechanism that produces bimodal execution durations and overshadows CPU effects.
We observe that the CPU composition on AWS Lambda shifts within a single week.
Additionally, we reveal that on Google Cloud Run Functions and Alibaba Function Compute, the availability of instruction set extensions such as SHA-NI differs between CPU types, widening the performance gap further.
Since all platforms bill by duration, performance gaps between CPU types translate directly into cost, with invocations on the slowest CPU type costing up to three times as much as those on the fastest.
\end{abstract}



\maketitle
\section{Introduction}
\label{sec:introduction}

\begin{figure}
    \centering
    \includegraphics[width=.8\linewidth]{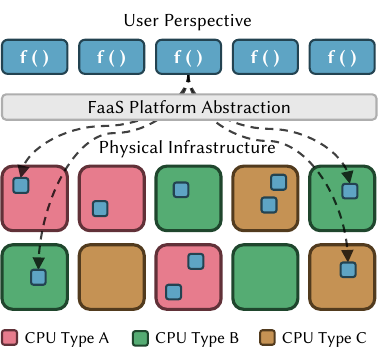}
    \caption{
        FaaS platforms abstract the underlying infrastructure behind a function interface.
        Requests to the same function may run on hosts with different CPU types, which users can neither see nor control.}
    \label{fig:problem}
\end{figure}
Function-as-a-Service (FaaS) has become an established cloud computing paradigm over the past decade.
Developers decompose applications into small, event-driven units of code, known as serverless functions.
The platform, in turn, manages the underlying execution environments, including resource provisioning, scaling, and load balancing, abstracting operational concerns from users~\cite{jonas2019berkeley,li2022survey}.
Platforms dynamically multiplex these execution environments, which are also referred to as function instances, across shared infrastructure, improving resource utilization.
Users benefit from a pay-per-use pricing model, only paying for the actual usage of their functions rather than for reserved or pre-allocated capacity~\cite{jonas2019berkeley}.

A direct consequence of this model is that users have limited control over the hardware their code runs on, as platforms expose only a small set of predefined resource configurations, such as the memory size~\cite{jonas2019berkeley}, rather than instance families as in Infrastructure-as-a-Service (IaaS)~\cite{aws2026ec2instances}.
This abstraction hides a heterogeneous CPU fleet underneath~\cite{wang2018peeking}, meaning that otherwise identical invocations may execute on different CPU types as instances scale out (see \cref{fig:problem}).
Lower-performing hardware consequently increases both execution duration and cost, as users are billed for execution duration.
The ephemeral nature of function instances~\cite{wang2018peeking,shahrad2020wild} exacerbates this effect.
They turn over more often than long-lived virtual machines in IaaS, potentially redrawing the hardware assignment with every new instance, which increases the performance variability between invocations.
Prior studies document the hardware heterogeneity in FaaS~\cite{wang2018peeking,figiela2018performance} and performance differences between CPU types~\cite{cordingly2020predicting}.
However, how much of the overall performance variability is attributable to hardware heterogeneity, and how this contribution differs across providers, remains unquantified.

Because commercial FaaS platforms reveal neither the hardware they assign nor the scheduling policies behind it, attributing performance differences to hardware requires measuring the platforms from the outside.
In this paper, we therefore conduct a systematic measurement study across four commercial platforms (AWS Lambda~\cite{aws2026lambda}, Azure Functions~\cite{microsoft2026flexconsumption}, Google Cloud Run Functions~\cite{google2026functions}, and Alibaba Function Compute~\cite{alibaba2026functioncompute}), characterizing their hardware assignment patterns and quantifying how much of the observed performance variability hardware heterogeneity explains.

We find that CPU type explains up to \qty{97}{\percent} of performance variance and can triple invocation cost for hardware-accelerated tasks, although on Azure other platform characteristics overshadow its effect.
We make the following contributions.
\begin{itemize}
    \item We design a platform-agnostic benchmark that samples the heterogeneous hardware underlying commercial FaaS platforms and captures function performance alongside CPU metadata (\cref{sec:benchmark_design}).
    \item We apply this benchmark to four commercial FaaS platforms, use variance decomposition to attribute the observed performance variability to hardware heterogeneity, and quantify the resulting cost differences from billed durations (\cref{sec:evaluation}).
    \item We discuss implications of our findings for users running production workloads and for researchers designing reproducible FaaS benchmarks (\cref{sec:discussion}).
\end{itemize}

\section{From Hardware to Serverless Performance}
\label{sec:background}
To save cost and increase resource utilization, serverless platforms use the heterogeneous hardware present in their datacenters to run function instances~\cite{jonas2019berkeley}.
These instances draw on the same pool of heterogeneous hardware as other cloud service instances, but with less user control over the resulting assignment.
This user-opaque assignment makes hardware heterogeneity an uncontrollable source of function performance variability.

\subsection{FaaS Key Characteristics}
In line with simplifying operational concerns for users, platforms expose only few options to configure the resources of a function instance~\cite{jonas2019berkeley}.
Often, only the memory size is configurable, and the platform derives the other resources proportionally~\cite{kuhlenkamp2022synthesizing}.
While users define these static parameters at deployment time, the runtime lifecycle of function instances is fully managed by the platform.
Upon function invocation, the platform either reuses an existing idle function instance, also known as a warm start, or provisions a new one, known as a cold start~\cite{shahrad2020wild,bermbach2020coldstarts}.
Whether to reuse, provision, or terminate an instance is entirely up to the platform.

Another defining characteristic of FaaS is its pay-per-use billing model, where users are only charged for the resources they actually consume~\cite{jonas2019berkeley}.
Specifically, billing is determined by the total execution duration of all function invocations scaled by the configured resources, plus a potential per-invocation fee.
Consequently, a function's execution duration becomes the primary driver of cost~\cite{lin2026demystifying}.

\subsection{Hardware Heterogeneity in Cloud Datacenters}
Cloud providers operate datacenters across multiple geographic regions, housing thousands of servers procured over many years.
As infrastructure continuously expands and new hardware generations are integrated, different generations of hardware concurrently operate within the same datacenter~\cite{barroso2026datacenter}.
Although operators employ virtualization to abstract this heterogeneity from tenants, the underlying hardware differences remain visible as performance variation to users.
The performance of the same workload on identical instances of the same cloud service can vary significantly depending on the underlying hardware~\cite{ou2013instance,schad2010runtime,lenk2011paying}.

This effect also extends to FaaS platforms, where function execution environments run on fleets of heterogeneous CPUs~\cite{wang2018peeking,figiela2018performance,cordingly2025sky}.
Unlike a long-running virtual machine in IaaS, which users provision once and which then typically runs on the same physical host for the lifetime of the deployment~\cite{ou2013instance}, FaaS instances are ephemeral and continually reprovisioned to serve fluctuating workloads~\cite{bermbach2021future}.

\subsection{Quantifying Performance Variability}
Performance variability in computer systems refers to the dispersion of a performance metric observed across multiple executions of the same workload under similar conditions~\cite{jain1991performance}.
In FaaS, this manifests as differences in execution duration across invocations of the same function under the same configuration, a variance that is substantial yet often overlooked in serverless research~\cite{wen2025variance}.
To quantify this variability, we use the coefficient of variation (CV), defined as the ratio of the standard deviation to the mean, which is dimensionless and thus comparable across platforms and configurations~\cite{jain1991performance}.

Quantifying the overall variability alone, however, is insufficient to understand its causes.
FaaS platforms experience variability from multiple sources, including cold starts, noisy neighbors, and configuration differences, all of which can confound the effect of hardware heterogeneity on performance.
Separating the contribution of hardware differences from these other sources requires appropriate statistical methods.
Effect size measures, specifically $\eta^2$, serve this purpose by expressing the proportion of variance in a response variable that a given factor accounts for~\cite{lakens2013effect}.
For a grouping factor $g$ that partitions the observations into groups, a one-way analysis of variance (ANOVA) decomposes the total sum of squares into a between-group and a within-group component, and $\eta^2$ is the ratio of the between-group sum of squares to the total,
\begin{equation}
\eta^2 = \frac{\mathrm{SS}_\mathrm{between}}{\mathrm{SS}_\mathrm{total}} = \frac{\sum_g n_g (\bar{y}_g - \bar{y})^2}{\sum_i (y_i - \bar{y})^2}
\label{eq:eta2}
\end{equation}
where $y_i$ are the individual observations, $\bar{y}_g$ the mean of group~$g$, $\bar{y}$ the grand mean, and $n_g$ the number of observations in group~$g$.
We later use this measure with CPU type as the grouping factor to quantify how much of the observed performance variability hardware heterogeneity accounts for~(\cref{sec:workload_sensitivity}).
The between-group component then captures the differences between CPU types, whereas all other sources, such as noisy neighbors, end up in the within-group component, which is how the decomposition separates them from hardware heterogeneity.

\section{Benchmark Design}
\label{sec:benchmark_design}
To explore the impact of hardware heterogeneity on performance variability in FaaS platforms, we introduce a systematic benchmarking methodology that runs microbenchmarks inside serverless functions and captures their performance alongside CPU metadata.
Our benchmark targets two primary dimensions in which hardware heterogeneity manifests, namely the composition of CPU types that function instances are scheduled on, and the extent to which performance differs between those CPU types.

Reliable measurements require considering how workloads interact with platform scheduling.
While placement is up to the platform, the invocation pattern of a benchmark influences how many instances it encounters and thus which hardware it samples.
Sequential invocations can be routed to the same function instance, repeatedly sampling the same hardware and limiting exposure to hardware heterogeneity.
Parallel requests, in contrast, force the platform to provision additional instances to serve the load, raising the likelihood of sampling different hardware.
A purely parallel approach, however, would mostly measure cold starts rather than steady-state performance.
We therefore generate a high volume of parallel invocations while allowing multiple sequential invocations per instance.
Beyond the invocation pattern, our benchmark comprises a set of platform-agnostic components that together enable comparison across platforms, regions, and resource configurations.
The source code is publicly available.\!\footnote{\url{https://github.com/Persists/benchmark-classifaas}}

\subsection{Function Design}
\label{sec:function_design}
The first component comprises the functions, which a client-side module deploys while abstracting provider-specific deployment and call signatures.
Each function is implemented for the Node.js~20 runtime, invoked via HTTPS, and consists of a wrapper and a microbenchmark.
In each invocation, the wrapper extracts hardware and instance metadata and then runs a CPU-intensive microbenchmark.
We measure function performance through the execution duration of the microbenchmark rather than the billed duration, since the billed duration also includes additional overhead such as initialization steps~\cite{lin2026demystifying}.

We employ five synthetic, compute-bound microbenchmarks designed to stress different CPU performance characteristics, chosen over application-level workloads to isolate CPU performance from confounding factors such as network latency and I/O~\cite{bermbach2017book}.
We use matrix multiplication as a baseline CPU-bound workload commonly employed in serverless benchmarking studies~\cite{werner2018matrix}.
JSON serialization repeatedly serializes and deserializes a deeply nested object structure, and gzip compression compresses a random buffer, both representing operations frequently found in serverless applications such as API request processing and payload transfer.
Finally, SHA-256 repeatedly hashes a large buffer and AES-CTR repeatedly encrypts a large buffer, both benefiting from the optional instruction set extensions SHA-NI and AES-NI, respectively, allowing us to quantify the impact of their availability on performance variability.

\subsection{Load Generation}
The second component, the load generator, runs client-side, drives function invocations, and collects performance measurements.
Motivated by the considerations above, it is designed to provoke the platform into provisioning new function instances.
To this end, it employs a closed-loop workload model in which a fixed number of concurrent workers continuously issue requests to the same function~\cite{schroeder2006open}.
We prefer a closed-loop model because, in an open-loop model with a fixed arrival rate, the number of concurrent instances would depend on the execution duration, and thus on the very CPU types we study.

\subsection{Instance Lifecycle Management}
\label{sec:instance_lifecycle_management}

Complementing the parallel invocation pattern of the load generator, we integrate a recycling mechanism to emulate workload patterns that exercise instance turnover.
Beyond the platform's scaling to serve concurrent requests, fresh instances are also provisioned in response to runtime crashes or other process failures~\cite{schirmer2025minos}.
To approximate this, we track the invocation count in a global variable within the runtime and terminate the process once it reaches a fixed threshold, prompting the platform to replace the terminated process before serving subsequent requests.
This approach also addresses concurrency limits enforced by platforms per region, function, or account, which cap the number of coexisting function instances.
How a platform responds to this turnover, however, is not under our control and may differ across platforms.
Either it provisions a fresh instance on a potentially different physical host or restarts the runtime process within the existing instance and continues serving subsequent invocations on the same hardware.
In our experiments, GCF and Alibaba Function Compute provision a fresh instance, whereas AWS Lambda and Azure Functions restart the runtime process within the existing instance, which we discuss in detail in \cref{sec:threats_to_validity}.

We set the recycling threshold to four invocations, yielding a maximum of four invocations per runtime process.
Since the first invocation of each runtime process includes initialization overhead, we discard it to capture steady-state performance rather than warm-up effects.
As every new function instance starts a new runtime process, this also removes instance cold-start effects.
Additionally, we trim the upper \qty{1}{\percent} of execution durations within each CPU type and configuration for the computation of outlier-sensitive metrics such as the CV and $\eta^2$, since skewed distributions otherwise distort these measures~\cite{jain1991performance}.
No lower trimming is required, since the microbenchmarks impose a natural floor on execution duration.

\subsection{Processor Identification}
As FaaS platforms do not expose information about the underlying hardware, our benchmark follows \citeauthor{cordingly2020saaf}~\cite{cordingly2020saaf} and extracts CPU metadata at runtime from within each function instance by parsing the \texttt{/proc/cpuinfo} file. The wrapper captures this metadata alongside a runtime process identifier and the execution duration of each invocation so that measurements can be grouped by process and CPU type later on.
The amount of information exposed through \texttt{/proc/cpuinfo} varies across platforms, consistent with prior work~\cite{wang2018peeking}.
We therefore construct CPU type identifiers using the most specific combination of available fields on each platform, resulting in different grouping granularities across platforms.
Coarser groupings, however, yield conservative estimates of the impact of CPU heterogeneity, since pooling distinct CPU models under one identifier moves their performance differences into the within-group variance rather than the between-group variance.
Moreover, the CPU type may co-vary with other host components that cannot be distinguished from within a function, such as the memory hierarchy, so we treat it as the observable identifier of the host configuration rather than of the processor alone.

\begin{figure*}[t]
    \centering
    \includegraphics{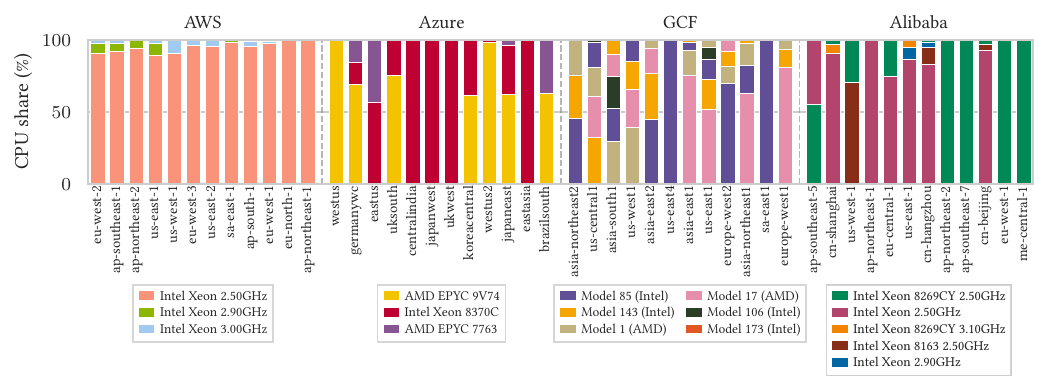}
    \caption{CPU type composition in 12 regions per platform, measured as the share of invocations.}
    \label{fig:composition_all_providers}
\end{figure*}

\section{Evaluation}
\label{sec:evaluation}
We use our benchmark to evaluate hardware heterogeneity and its contribution to performance variability across four commercial FaaS platforms, with the results publicly available.\!\footnote{\url{https://github.com/Persists/results-classifaas}}
We structure our investigation into three sets of experiments, covering regional differences~(\cref{sec:regional_diversity}), configurational effects~(\cref{configurational_effects}), and temporal dynamics over seven days~(\cref{sec:temporal_dynamics}).
We then use the same seven-day dataset to analyze how different workloads react to these hardware differences~(\cref{sec:workload_sensitivity}) and how the resulting performance differences translate into cost~(\cref{sec:cost_implications}).

\subsection{Platforms and Configuration}
\label{sec:experimental_setup}
We leverage our benchmark to study AWS Lambda~\cite{aws2026lambda}, Google Cloud Run Functions (GCF)~\cite{google2026functions}, Azure Functions~\cite{microsoft2026flexconsumption}, and Alibaba Function Compute~\cite{alibaba2026functioncompute}, which together provide broad market coverage and enable meaningful cross-platform comparison.
Since several platforms offer multiple hosting plans or product generations with distinct performance characteristics, we fix one variant per platform for all experiments, namely Gen~1 on GCF and the Flex Consumption plan on Azure Functions, a Linux-based option that exposes explicit memory configuration and supports horizontal scaling.
Azure Functions and Alibaba Function Compute allow multiple concurrent invocations per function instance by default, which introduces performance interference between requests sharing the same instance~\cite{lin2026demystifying}.
We therefore set per-instance concurrency to one on both, matching the one-request-per-instance default of AWS Lambda and GCF.
Alibaba is the only platform that decouples vCPU from memory size.
To aid comparability between platforms, we configure it with \qty{0.1}{vCPU} per \qty{128}{\mega\byte} of memory.
Finally, we calibrate the input parameters of each microbenchmark to yield an execution duration of approximately \qty{2}{\second} on AWS Lambda at \qty{512}{\mega\byte}.

\subsection{Regional Differences in CPU Fleets}
\label{sec:regional_diversity}

To gain an initial understanding of the hardware heterogeneity landscape, we start by characterizing the CPU types present across different regions on each platform.
We deploy the same matrix multiplication function in 12 randomly selected regions per provider with a fixed memory configuration of \qty{512}{\mega\byte}, which is the lowest memory size offered by all four platforms.
We schedule a separate load generator for each region that invokes the corresponding function \num{2500} times in total.
Each generator issues 500 parallel invocations on AWS Lambda, GCF, and Alibaba Function Compute, and 300 on Azure Functions, the highest count the platform handled reliably in our setup.
To mitigate temporal performance variations present on serverless platforms~\cite{schirmer2023nightshift}, we schedule all regional experiments on November 8, 2025, to start at the same local time in every region.

The resulting CPU type distributions are shown in \cref{fig:composition_all_providers}.
Azure reports the full model name for all CPUs, whereas AWS reports a generalized identifier consisting of vendor and clock frequency, and Alibaba falls in between, reporting the full model name for some CPUs and a generalized identifier for others.
GCF masks the model name as ``unknown'', matching prior findings from \citeauthor{wang2018peeking}~\cite{wang2018peeking}.
For GCF, we therefore group invocations by the combination of the reported vendor (AMD or Intel) and CPUID model number, a labeling scheme we retain throughout the remaining experiments.

Every provider employs heterogeneous CPU types within and across regions, but the degree and distribution patterns vary substantially.
AWS exhibits the most homogeneous fleet, consisting of only three CPU types, with most regions dominated by a single Intel Xeon \qty{2.50}{\giga\hertz} type that accounts for over \qty{90}{\percent} of invocations.
In contrast, GCF shows the most heterogeneous fleet overall: only two regions are dominated by Model~85, while the others are distributed more evenly across multiple types.
Azure and Alibaba fall in between, with some regions served exclusively by a single CPU type and others exhibiting a more mixed composition.
Across the full sample, GCF and Alibaba expose the most distinct CPU types, though later experiments reveal that AWS deploys additional types not captured in this snapshot.

\begin{figure}
    \centering
    \includegraphics{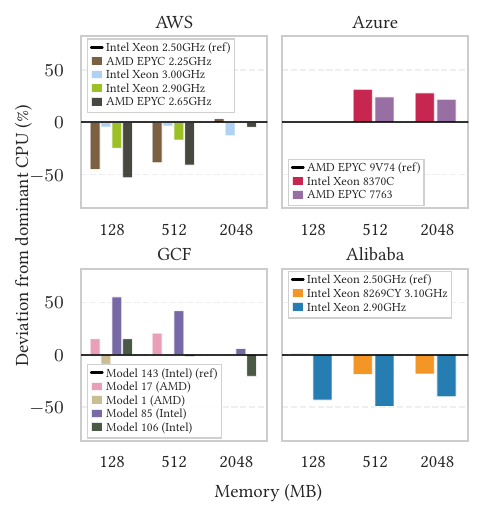}
    \caption{Deviation of the median execution duration from each platform's dominant CPU type (ref) across memory tiers.}
    \label{fig:relative_performance_deviation}
\end{figure}

\subsection{Configurational Effects}
\label{configurational_effects}
Having established that CPU heterogeneity exists across and within regions, we now examine how function configuration affects hardware assignment.
We select a region with a diverse CPU composition in the regional survey for each platform, namely \texttt{us-east-1} for AWS Lambda and Alibaba Function Compute, \texttt{us-east1} for GCF, and \texttt{germanywestcentral} for Azure Functions.
In each region, we deploy the matrix multiplication benchmark at three memory tiers (\qtylist[list-units=single]{128;512;2048}{\mega\byte}), except on Azure, where the Flex Consumption plan only offers \qtylist[list-units=single]{512;2048}{\mega\byte} from this range.
We repeat the experiment 12 times at \qty{2}{\hour} intervals on November 26--27, 2025, covering a full \qty{24}{\hour} cycle, and invoke all functions simultaneously within each run to minimize temporal effects between configurations.
Each configuration of the same function receives \num{1500} invocations per run, with \num{250} concurrent invocations so that the simultaneously running configurations stay within each platform's concurrency limits, for a total of \num{18000} invocations per configuration.

\subsubsection{CPU Composition}
The CPU composition remains stable across memory configurations on all four platforms, with the dominant CPU types unchanged and only minor fluctuations in their shares.
AWS Lambda's fleet in \texttt{us-east-1} shows two CPU types absent from the regional survey, the AMD EPYC \qty{2.25}{\giga\hertz} and the AMD EPYC \qty{2.65}{\giga\hertz}, which together account for more than \qty{16}{\percent} of invocations, a difference we examine in \cref{sec:temporal_dynamics}.

\subsubsection{Relative Performance Differences Across Configurations}

AWS Lambda and GCF show pronounced inter-CPU differences at low memory tiers that diminish with increasing memory, while Azure and Alibaba maintain stable differences across all tiers (\cref{fig:relative_performance_deviation}).
On AWS Lambda, the median execution duration of the AMD EPYC \qty{2.25}{\giga\hertz} is \qty{45}{\percent} lower than the dominant Intel Xeon \qty{2.50}{\giga\hertz} at \qty{128}{\mega\byte} but \qty{4}{\percent} higher at \qty{2048}{\mega\byte}.
GCF shows the same convergence, most pronounced for Model~85, which yields \qty{55}{\percent} higher execution durations than the dominant Model~143 at \qty{128}{\mega\byte} but only \qty{6}{\percent} higher at \qty{2048}{\mega\byte}.
For users deploying on these platforms, this means that performance variability due to CPU heterogeneity is most pronounced at low memory tiers, while higher tiers offer more predictable performance regardless of the assigned CPU type.

We hypothesize that this convergence reflects context-switch penalties that differ across CPU architectures~\cite{mogul1991context}.
Lower memory tiers receive a smaller share of a CPU core that is time-sliced among function instances.
Consequently, the same work takes longer, the process is descheduled more often, and co-located instances pollute the cache in the meantime.
Each context switch thus leaves the process with a partially cold cache, and since CPU architectures recover from this at different speeds, the penalty widens their performance gap at low tiers, whereas at high tiers, with few switches per invocation, the gap shrinks to the difference in raw compute speed.
AWS Lambda is the only platform that exposes the context switch counter in \texttt{/proc/stat}, which we record inside the execution environment before and after each invocation.
The number of context switches per invocation falls by a factor of \numrange{3.7}{9.0} from \qty{128}{\mega\byte} to \qty{2048}{\mega\byte}, depending on the CPU type, consistent with this explanation.

On Alibaba, the Intel Xeon \qty{2.90}{\giga\hertz} consistently achieves \qtyrange{40}{50}{\percent} lower execution durations than the dominant Intel Xeon \qty{2.50}{\giga\hertz}, and on Azure the other two CPU types take \qtyrange{22}{32}{\percent} longer than the dominant AMD EPYC 9V74.
Beyond these inter-CPU differences, Azure's overall variability at \qty{512}{\mega\byte} stands out with a CV of \qty{54}{\percent}, against only \qty{18}{\percent} at \qty{2048}{\mega\byte}.
Since both the inter-CPU differences and the CPU composition remain nearly identical across the two configurations, CPU heterogeneity cannot explain this gap, pointing to a separate driver of variability, which we examine next.

\subsubsection{Azure Bimodality}
\label{sec:azure_bimodality}
\begin{figure}
    \centering
    \includegraphics{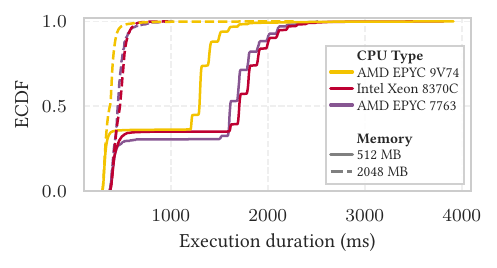}
    \caption{Empirical CDFs of execution durations on Azure Functions. At \qty{512}{\mega\byte}, every CPU type is bimodal, with a subset of invocations as fast as at \qty{2048}{\mega\byte}.}
    \label{fig:azure_bimodality}
\end{figure}

Closer inspection of the \qty{512}{\mega\byte} execution durations in \cref{fig:azure_bimodality} reveals a bimodal distribution, with up to \qty{36}{\percent} of invocations completing roughly \qty{4}{\times} faster than the median.
This bimodality is present across all three CPU types, suggesting that a platform-level mechanism rather than a CPU-specific effect drives the performance variability.
We observe the same pattern across every region evaluated in \cref{sec:regional_diversity}, though its extent varies across regions, with some exhibiting a substantially larger share of faster invocations than others.

Notably, on each CPU type, the faster \qty{512}{\mega\byte} invocations align with the \qty{2048}{\mega\byte} performance.
This suggests that a subset of \qty{512}{\mega\byte} instances may receive a higher effective compute allocation than expected from the nominal resource specification. 
According to the Azure Flex Consumption documentation, the \qty{512}{\mega\byte} tier typically provides \qty{0.25}{vCPU}, whereas the \qty{2048}{\mega\byte} tier provides \qty{1}{vCPU}. 
The documentation further states that initial instances may receive different core allocations to improve performance, which is consistent with the observed execution duration distributions~\cite{microsoft2026flexconsumption}.
In contrast, the \qty{2048}{\mega\byte} tier exhibits a single tightly clustered distribution without visible bimodality.

\subsection{Temporal Dynamics of CPU Assignment}
\label{sec:temporal_dynamics}
\begin{figure}
    \centering
    \includegraphics{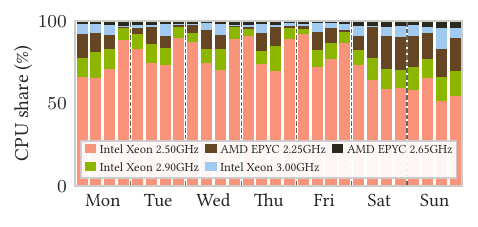}
    \caption{CPU type composition on AWS Lambda over a seven-day period, aggregated into four six-hour time slots (00:00, 06:00, 12:00, 18:00) per day in the region's local time.}
    \label{fig:aws_composition_over_time}
\end{figure}

While the previous experiments provide snapshots at specific points in time using a single workload, this final set of experiments investigates a broader temporal profile across a full seven-day period (December 1--7, 2025).
We deploy each of the five microbenchmarks~(\cref{sec:function_design}) as a separate function with \qty{512}{\mega\byte} of memory in the same regions as in \cref{configurational_effects}.
Across the entire period, we schedule the experiments in four daily time slots at 00:00, 06:00, 12:00, and 18:00 in the local time of each region's datacenter, to capture both diurnal and day-to-day variability.
To stay within platform-imposed concurrency limits, we execute the microbenchmarks sequentially within each time slot, with a \qty{10}{\minute} delay between runs.
Furthermore, we randomize the execution order of the workloads within each time slot following the randomized multiple interleaved trials (RMIT) approach~\cite{abedi2017repeatable}, reducing bias from ordering effects to keep the workloads comparable for the analyses in \cref{sec:workload_sensitivity,sec:cost_implications}. 
Each run uses the same settings as the regional survey~(\cref{sec:regional_diversity}), \num{2500} invocations with \num{300} to \num{500} parallel invocations depending on the platform, totaling \num{70000} invocations per workload and platform over the seven days.

To investigate the temporal dynamics of CPU compositions, we pool the CPU shares of all five microbenchmarks per time slot.
Since every slot contains the same five microbenchmarks, changes between slots reflect the CPU fleet rather than differences between workloads.
The results show that Azure, GCF, and Alibaba maintain stable CPU compositions throughout the week, with the dominant CPU types and their shares remaining nearly identical over the entire period.
AWS Lambda, in contrast, shows pronounced temporal shifts in its CPU composition, as highlighted in \cref{fig:aws_composition_over_time}.
While the Intel Xeon \qty{2.50}{\giga\hertz} remains the dominant type over the full period, its share fluctuates within and across days.
On weekdays, the two middle time slots (06:00 and 12:00) show a higher share of the faster AMD and Intel variants, whereas the Intel Xeon \qty{2.50}{\giga\hertz} takes a larger share in the outer slots (00:00 and 18:00).
On weekends, the faster variants take a larger share across all time slots.

These diurnal and weekly shifts, however, do not fully explain the composition discrepancy between the prior experiments in \cref{sec:regional_diversity} and \cref{configurational_effects}.
The regional survey encountered no AMD CPU in \texttt{us-east-1}, although it ran on a Saturday morning, a time slot in which AMD CPUs are well represented in our seven-day experiment.
Beyond the recurring patterns, the fleet composition thus also changed in the weeks between the experiments.

\subsection{Workload Sensitivity to CPU Heterogeneity}
\label{sec:workload_sensitivity}

\begin{figure}
    \centering
    \includegraphics{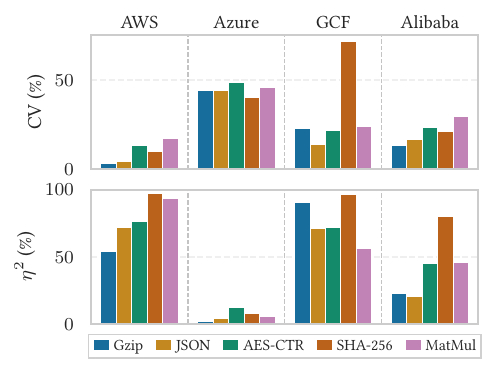}
    \caption{Coefficient of variation and the share of variance explained by CPU type ($\eta^2$) for each provider and microbenchmark at \qty{512}{\mega\byte}.}
    \label{fig:cv_eta2_combined}
\end{figure}

The seven-day dataset across five microbenchmarks additionally allows us to quantify how much CPU type contributes to the overall variability of each workload.
To this end, \cref{fig:cv_eta2_combined} reports two complementary measures, the CV for total observed variability and $\eta^2$~(\cref{eq:eta2}) for the share of that variability attributable to CPU type.
We compute $\eta^2$ for each platform and workload from a one-way ANOVA with CPU type as the grouping factor on log-transformed execution durations.
The log scale accounts for the multiplicative nature of inter-CPU performance differences and for the right-skewed execution duration distributions, following standard practice for performance measurements~\cite{jain1991performance}.

Two regimes emerge.
On AWS Lambda and GCF, $\eta^2$ ranges from \qtyrange{54}{97}{\percent} and \qtyrange{56}{96}{\percent}, respectively.
Hence, on these platforms most of the observed performance variability can be explained by differences in the underlying CPU type, consistent with narrow distributions within each CPU type, where the per-CPU CV stays below \qty{11}{\percent} for every workload on AWS Lambda and below \qty{21}{\percent} on GCF.
Although AWS Lambda has the lowest CVs of any platform across all workloads, even its small residual variability is largely attributable to the assigned CPU type.

Azure Functions shows the opposite pattern.
Its CVs are consistently high, yet CPU type explains at most \qty{12}{\percent} of the variance, because the bimodality identified in \cref{sec:azure_bimodality} appears for every workload.
Alibaba Function Compute sits between these two regimes, with $\eta^2$ varying substantially across workloads, from \qty{21}{\percent} on JSON to \qty{80}{\percent} on SHA-256, as its distributions within each CPU type are considerably wider, with per-CPU CVs of up to \qty{39}{\percent}.

SHA-256 yields the highest $\eta^2$ on all platforms except Azure Functions, reaching \qty{97}{\percent} on AWS Lambda, \qty{96}{\percent} on GCF, and \qty{80}{\percent} on Alibaba.
On GCF and Alibaba, this traces back to instruction set extensions.
The CPU flags in \texttt{/proc/cpuinfo} show that Model~85 is the only CPU type on GCF without SHA-NI, and it takes \qtyrange{3.1}{5.3}{\times} as long as the other types, also explaining the high CV on GCF.
On Alibaba, conversely, the Intel Xeon \qty{2.90}{\giga\hertz} is the only CPU type with SHA-NI support and is \qtyrange{3.5}{4.2}{\times} faster than the others.
On AWS Lambda, no CPU type exposes SHA-NI, and on Azure Functions all of them do, so neither platform exhibits such inter-CPU differences for SHA-256.

\subsection{Cost Implications of CPU Assignment}
\label{sec:cost_implications}
\begin{figure}
    \centering
    \includegraphics{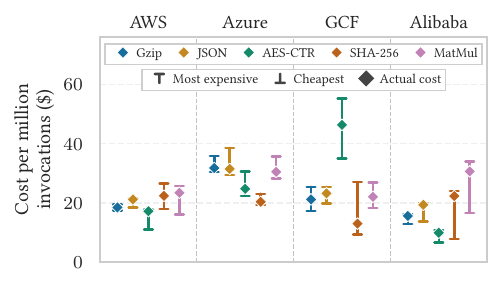}
    \caption{
        Mean cost per million invocations, derived from billed durations at list prices including the per-request fee.
    }
    \label{fig:cost_per_cpu}
\end{figure}

The execution duration differences characterized so far are not only a performance concern.
All four platforms charge for the duration of an invocation at a rate that does not depend on the assigned CPU type, so an invocation that lands on a slower CPU also incurs a higher cost.
Therefore, we now derive cost for the seven-day dataset from the platform-reported billed durations at the pay-as-you-go list prices without free tiers~\cite{aws2026lambdapricing,microsoft2026functionspricing,google2026functionspricing,alibaba2026functionpricing}, applying each platform's billing granularity, including the minimum of \qty{1000}{\milli\second} on Azure Functions~\cite{microsoft2026flexconsumption}.
However, while every invocation performs the same amount of work, the billed duration also covers the metadata collection of our wrapper, and each invocation incurs a fixed request fee.
Both compress relative differences, so the cost differences we report are slightly smaller than they would be for a function without our metadata collection.
Finally, we calculate the mean cost per CPU type over one million invocations.

\Cref{fig:cost_per_cpu} shows the resulting cost of invocations running on the cheapest and the most expensive CPU type, considering only types with a share of at least \qty{1}{\percent}, together with the actual cost, i.e., the average over all invocations given the CPU types the platform assigned.
Overall, the differences are smallest on AWS Lambda and Azure Functions, where invocations on the most expensive CPU type still cost up to \qty{1.6}{\times} and \qty{1.4}{\times} as much as those on the cheapest one, respectively.
They are largest on GCF and Alibaba Function Compute, with up to \qty{2.9}{\times} and \qty{3.0}{\times} for SHA-256, where only some CPU types support SHA-NI, consistent with the findings in \cref{sec:workload_sensitivity}.
Where the actual cost falls within this range depends on how often each CPU type is assigned.
On Alibaba Function Compute, it is close to the most expensive CPU type for all workloads.
AWS Lambda shows a similar pattern for JSON, AES-CTR, and MatMul, while the remaining workloads fall towards the middle of the range.
GCF also falls around the middle, except for SHA-256, where the actual cost is close to the cheapest type because the CPU type without SHA-NI support is rarely assigned.
In contrast, Azure Functions stays close to the cheapest type for every workload.

\section{Discussion}
\label{sec:discussion}
Our results show that the underlying hardware, which FaaS platforms promise to abstract away, remains a primary driver of user-facing performance variability.
We find that CPU type explains \qtyrange{54}{97}{\percent} of the performance variance on AWS Lambda and GCF.
While the serverless abstraction holds at the interface, it breaks down in the resulting performance, exposing users to this hidden hardware lottery.
The Azure results, however, show that hardware is not a universal explanation for FaaS variability, since a platform mechanism overshadows CPU effects.

\subsection{Users Facing the Hardware Lottery}
\label{sec:implications_users}
For users, hardware-dependent variability is systematic rather than transient.
Once a function instance is assigned to a physical host, all of its invocations inherit that host's performance and, due to pay-per-use billing, its cost, until the platform recycles the instance.
This distinguishes hardware-dependent variability from noisy neighbors or network jitter, which can fluctuate within an instance's lifetime.
The CPU type can, however, be read at runtime at negligible cost, and logging it alongside latency and billing data lets users attribute anomalies to hardware rather than to their code.
Whether a slowdown stems from the processor itself or from co-varying components such as the memory hierarchy is irrelevant for this purpose, since the CPU type is the observable property that predicts both performance and cost.
On AWS Lambda and GCF, the differences between CPU types shrink with increasing memory, so functions with strict latency requirements can gain predictability from a higher tier, even if they do not use the additional memory.
Workloads that rely on instruction set extensions such as SHA-NI face the largest gaps and should check the CPU flags at runtime rather than assume availability.

Beyond observation, users can attempt placement gaming to reduce cost or improve performance~\cite{ginzburg2020serverless,schirmer2025minos}, provoking fresh instances until a favorable CPU type is assigned.
Even if every invocation landed on the cheapest CPU type, the saving is bounded by the distance between the observed mix and the cheapest type in \cref{fig:cost_per_cpu}, which depends on both platform and workload and therefore has to be monitored per function.
At the same time, each attempt incurs a cold start, which adds latency and billed initialization time~\cite{lin2026demystifying}, and since the fleet on AWS Lambda shifts within days, current compositions should not be assumed to persist.
CPU identification is therefore more reliable for monitoring performance than for controlling placement.

\subsection{Scheduling Flexibility at the Cost of User Predictability}
\label{sec:implications_platforms}
Our results indicate that, on AWS Lambda and GCF, allowing users to select a CPU type could substantially reduce performance variability for their functions.
Such control is not unprecedented, as IaaS offerings let users select instance families~\cite{aws2026ec2instances} or a minimum CPU platform~\cite{google2026cpuplatforms}, and AWS Lambda already prices ARM and x86 functions differently~\cite{aws2026lambdapricing}.
Following this precedent, pricing could depend on the selected CPU type in addition to the memory tier.
For FaaS providers, however, CPU-dependent placement would reduce scheduling flexibility.
The current model of billing the entire fleet at a uniform rate allows providers to place instances on whichever compatible host is available, supporting high utilization through statistical multiplexing~\cite{jonas2019berkeley}.
Restricting placement to specific CPU types would fragment this pool and could increase the risk of underutilization.

We therefore propose three possible measures, in increasing order of cost to providers. 
First, making the assigned CPU type available as instance metadata would make the hardware lottery visible to users.
Second, normalizing billing by CPU performance could prevent users from paying more for the same workload on slower hardware.
Billing by consumed CPU time instead of wall-clock time, as proposed by \citeauthor{cao2025leopard}~\cite{cao2025leopard}, would not prevent this issue, since the same workload still consumes more CPU time on a slower CPU.
Lastly, an optional tier tied to a specific CPU generation, which may incur an additional cost, could allow users to exchange scheduling flexibility for predictability, while maintaining the current model for other workloads.

\subsection{Hardware Awareness as a Serverless Benchmarking Requirement}
\label{sec:implications_researchers}
In contrast to users, who value the ease of use of FaaS and might accept the hardware lottery as a trade-off, researchers require reproducibility and comparability in their experiments~\cite{bermbach2017book}.
CPU assignment is an uncontrolled experimental variable, as two experiments with identical function code and resource configurations may encounter different CPU compositions and therefore differ for reasons unrelated to the factor under investigation.
Beyond known time-of-day effects~\cite{schirmer2023nightshift}, fleet composition itself shifts over time and differs across regions. 
Region and measurement period are therefore part of the experimental context, an issue \citeauthor{schad2010runtime}~\cite{schad2010runtime} already raised for IaaS.

We therefore recommend concrete measures to mitigate this confounding factor in future FaaS benchmarking studies. 
Researchers should record the CPU type of every function instance, which is available at runtime with negligible overhead. 
They should then stratify the results by CPU type, or at least verify that the conditions being compared encountered a similar CPU composition, before attributing a difference to the factor under investigation.
Researchers should also ensure that they sample many instances rather than a few instances repeated many times, since hardware is assigned per instance and repeated invocations of one instance only resample the same hardware.
Finally, reporting the CPU composition renders this confounding factor explicit, enabling subsequent studies to evaluate whether their measurements were obtained on comparable hardware.

\subsection{Threats to Validity}
\label{sec:threats_to_validity}
We restricted our analysis to JavaScript functions and a fixed set of CPU-bound microbenchmarks.
This simplification strengthens our attribution of variance to CPU type, but it also means that the magnitude of variability for other languages, runtimes, and application-level workloads may differ from our reported values.
Moreover, our evaluation focuses on the variance associated with the observable CPU type rather than an isolated causal effect of the processor, as the CPU type may co-vary with other components of the host, such as the memory hierarchy or rack-level hardware configuration.
Additionally, the regions we selected for the configuration and temporal experiments were chosen for their heterogeneity in the initial regional survey, which means our results characterize regions where heterogeneity is present rather than the platform as a whole.

The recycling mechanism (\cref{sec:instance_lifecycle_management}) was only effective on GCF and Alibaba Function Compute.
We verify the instance identity through an identifier that the first invocation writes to the local \texttt{/tmp} filesystem, which persists even after process termination.
On GCF and Alibaba, a new identifier appears after almost every recycling, confirming a fresh instance, whereas on AWS Lambda and Azure Functions it persists, meaning that the runtime process was restarted within the existing instance.
On these two platforms, instances therefore serve more than four invocations, and our sampling reflects organic scaling under sustained parallel load rather than forced churn.
As a result, we sampled fewer distinct instances than intended.

Finally, CV and $\eta^2$ remain sensitive to skewness and heavy upper tails, which we mitigate by trimming the upper \qty{1}{\percent} of execution durations and by analyzing on the log scale for the decomposition.

\section{Related Work}
\label{sec:related_work}
The opacity of commercial serverless platforms has led to extensive efforts to infer platform characteristics solely through external measurements and observations.
Various studies have proposed benchmarking methodologies that stress different aspects of serverless platforms to gather insights about performance and other system properties~\cite{copik2021sebs, kim2019functionbench,yu2020serverlessbench,wen2023characterizing}, enabling cross-platform comparisons.
By combining benchmarks with controlled configurations, prior work has isolated specific performance factors within serverless platforms, such as the impact of CPU architecture in AWS Lambda when comparing x86 and ARM~\cite{lambion2022characterizing,chen2023x86}, the effect of resource configuration~\cite{figiela2018performance, lloyd2018factors} and programming language~\cite{cordingly2020language, manner2018coldstart}, or temporal effects derived from repeated measurements~\cite{schirmer2023nightshift,eismann2021review}.

While the explicit selection of available configurations or time windows allows for controlled setups, another line of work infers characteristics by leveraging the little information commercial FaaS platforms do not hide from users.
Several studies, e.g.,~\cite{wang2018peeking,figiela2018performance,cordingly2025sky}, extracted CPU metadata from \texttt{/proc/cpuinfo} to characterize the employed hardware.
\citeauthor{lloyd2018factors}~\cite{lloyd2018factors} investigated AWS Lambda by creating a temporary file on the local filesystem; if the file was already present during a subsequent invocation, they inferred that the same execution environment was reused, enabling analysis of cold-start behavior.
Utilizing the CPU information further, \citeauthor{awwad2025carbon}~\cite{awwad2025carbon} derived infrastructure sustainability properties such as power draw and operational and embodied carbon.
\citeauthor{lin2026demystifying}~\cite{lin2026demystifying} used user-space timing measurements to infer OS scheduler parameters and billing quantization effects on AWS Lambda, GCF, and IBM Cloud Functions.

Among inferred cloud platform characteristics, hardware heterogeneity has emerged as a significant driver of performance variability.
Over a decade ago, \citeauthor{schad2010runtime}~\cite{schad2010runtime} reported up to \qty{24}{\percent} CPU-performance variation on AWS EC2, primarily attributing it to hardware differences.
\citeauthor{lenk2011paying}~\cite{lenk2011paying} similarly traced a bimodal performance distribution on AWS EC2 to the assigned CPU type after ruling out datacenter location and time of day, and reported the probability of landing on each CPU type for a given instance class.
Later, these findings were revisited, revealing that the impact of hardware heterogeneity on performance had decreased~\cite{leitner2016patterns}.
In the serverless domain, the issue resurfaced with \citeauthor{cordingly2020predicting}~\cite{cordingly2020predicting} observing that CPU heterogeneity increased the CV significantly on AWS Lambda \((7.4\times)\), while the effect was considerably smaller on IBM Cloud Functions \((1.2\times)\), when comparing overall variance to the variance of a single CPU model.
During the evaluation of a serverless MapReduce framework, \citeauthor{gimenezalventosa2019framework}~\cite{gimenezalventosa2019framework} measured execution duration differences of up to \qty{51}{\percent} between CPU types on AWS Lambda.

Building upon inferred platform knowledge, different systems have been proposed to improve performance by exploiting favorable placements.
Even before serverless platforms emerged, \citeauthor{ou2013instance}~\cite{ou2013instance} exploited CPU heterogeneity within a single IaaS instance type on EC2 through a trial-and-better strategy that keeps fast instances and discards slow ones, achieving up to \qty{30}{\percent} cost savings, and analyzed how this benefit erodes once many users adopt it.
\citeauthor{ginzburg2020serverless}~\cite{ginzburg2020serverless} proposed placement-gaming strategies based on execution duration and CPU-steal measurements.
Similarly, Minos~\cite{schirmer2025minos} uses preliminary benchmarks to identify and selectively terminate slow instances, thereby achieving the same goal of favoring faster hardware.
\citeauthor{cordingly2025sky}~\cite{cordingly2025sky} present a sky computing approach that exploits CPU fleet differences across availability zones to redeploy functions to better-performing hardware.
To determine the CPU composition, they exhausted entire availability zones through repeated function invocations until the zone was saturated, exceeding a failure threshold. 

In contrast to prior work, which demonstrates that FaaS hardware is heterogeneous or exploits favorable placements, our work quantifies how strongly CPU type contributes to overall performance variability across four commercial FaaS platforms.
Rather than treating hardware heterogeneity as an isolated phenomenon, we decompose total variability to identify when hardware-induced effects dominate execution duration variability and when other factors become the primary source of variance.

\section{Conclusion}
\label{sec:conclusion}
In this paper, we showed that hardware heterogeneity is a first-order driver of performance variability in commercial FaaS platforms that users and researchers should account for when designing applications and evaluating platforms.
Combining compute-bound microbenchmarks with runtime hardware introspection, we quantified the contribution of CPU type to the observed variance on four platforms.
We found that each platform employs heterogeneous CPU fleets, and that on AWS Lambda these fleets shift within a single week rather than remaining static.
CPU type explained \qtyrange{54}{97}{\percent} of the variance in execution duration on AWS Lambda and GCF, but at most \qty{12}{\percent} on Azure Functions, where a platform-level mechanism produced bimodal execution durations.
Alibaba Function Compute sat between these regimes.

The common thread is that the serverless abstraction hides a source of variability it does not remove.
On AWS Lambda and GCF the hidden factor is the assigned CPU, while on Azure Functions it is an opaque platform mechanism, and in both cases users are exposed to performance and cost differences they can observe but not control.
This places workloads with tight performance or cost requirements in tension with the operational convenience that motivates the FaaS model.

\balance

\bibliographystyle{ACM-Reference-Format}
\bibliography{bibliography}

\end{document}